\documentclass[11pt]{article}

\usepackage{amsmath}

\begin{document}

\title{ V.M. Red'kov\footnote{redkov@dragon.bas-net.by}, E.M. Ovsiyuk\footnote{e.ovsiyuk@mail.ru},
O.V. Veko\footnote{vekoolga@mail.ru} \\
 Spin 1/2 particle   in the field of the Dirac string on the background of de Sitter
 space-time\\[3mm]
{\small Institute of Physics,  National Academy of Sciences of
Belarus
\\
Mosyr State Pedagogical University, Belarus}}

\maketitle

\begin{quotation}

The Dirac monopole string is specified for de Sitter
cosmolo\-gical model. Dirac equation  for spin 1/2 particle in
presence of this monopole has been examined on the background of
de Sitter space-time in static coordinates. Instead of spinor
monopole har\-monics, the technique of Wigner $D$-functions is
used. After separa\-tion of the variables, detailed analysis of
the  radial equations is performed; four types of solutions,
singular, regular, in- and out-  running waves,  are  constructed
in terms of hypergeometric functions. The complete  set of  spinor
wave solutions  $\Psi_{\epsilon, j,m, \lambda}(t,r, \theta, \phi)$
has
 been constructed, special  attention is given to treating  the states of minimal values of the  total
 angular moment  $j_{\min}$.

\end{quotation}


\section{ Introduction}

De Sitter and anti de Sitter geometrical models are given  steady attention in the context
of developing quantum theory in  a curved space-time  -- for instance, see in [1,2].
In particular, the problem of description  of the particles with different spins  on these curved backgrounds
 has a long history
-- see  [3--36].
Here we will be interested mostly in treating the Dirac equation in  de Sitter model.

In the present paper,  the influence of the  Dirac monopole string  on the spin 1/2 particle
in de Sitter cosmological model
is investigated\footnote{Such a
problem for spinless particle in the flat Minkowski space  was first considered by Dirac \cite{Dirac-1931}
 and Tamm \cite{Tamm-1931};
Harish-Chandra  \cite{Harish-Chandra} obtained the exact solution of Dirac equation for electron interacting
with magnetic-monopole field.}.
  Instead of spinor monopole harmonics, the technique of Wigner
$D$-functions is used. After separation of the variables radial equation have been solved
exactly in terms of hypergeometric functions.
The complete  set of  spinor wave solutions  $\Psi_{\epsilon, j,m, \lambda}(t,r, \theta, \phi)$  has
 been constructed. Special attention is given to treating  the states of minimal values for total
 angular momentum quantum number $j_{\min}$, these states turn to be much more complicated than in the flat Minkowski space.

\section{ Dirac particle in de Sitter  space}

The Dirac equation (the notation according  to \cite{Book-1} is
used)
\begin{eqnarray}
\left [ i \gamma^{c} \; ( e_{(c)}^{\alpha} \partial_{\alpha} + {1
\over 2} \sigma^{ab} \gamma_{abc}) - M \right  ]  \Psi = 0
\label{Dirac}
\end{eqnarray}

\noindent in static coordinates and tetrad of the Sitter space
\begin{eqnarray}
 dS^{2} = \Phi \;  dt^{2} - {dr^{2} \over \Phi } -
r^{2} (d\theta^{2} +  \sin ^{2}\theta d\phi ^{2}) \; ,  \; \; \Phi
=  1 - r^{2}  \; , \nonumber
\\
e^{\alpha}_{(0)}=({1 \over  \sqrt{\Phi} }, 0, 0, 0) \; , \qquad
e^{\alpha}_{(3)}=(0, \sqrt{\Phi}, 0, 0) \; , \nonumber
\\
e^{\alpha}_{(1)}=(0, 0, \frac {1}{r}, 0) \; , \qquad
e^{\alpha}_{(2)}=(1, 0, 0, \frac{1}{ r  \sin \theta})  \; ,
\nonumber
\\
\gamma_{030} ={ \Phi ' \over 2\sqrt{ \Phi }} \; , \; \gamma_{311}
={ \sqrt{ \Phi } \over r} \; , \; \gamma_{322} ={ \sqrt{ \Phi }
\over r} \; , \; \gamma_{122} ={ \cos \theta \over r \sin \theta}
\; ,
\end{eqnarray}

\noindent
takes the form
\begin{eqnarray}
\left [  i {\gamma ^{0}  \over \sqrt{\Phi}} \partial _{t}  + i
\sqrt{\Phi } \left (     \gamma ^{3}
\partial _{r}  + { \gamma ^{1} \sigma^{31}  +
\gamma  ^{2} j^{32} \over r }   \right. \right.
\nonumber
\\
\left. \left.   +{ \Phi' \over 2 \Phi  } \gamma ^{0}     \sigma^{03} \right  )  +
 { 1 \over r} \Sigma _{\theta,\phi }  -  M
 \right  ]    \Psi (x)  =  0 \; ,
 \label{10.1a}
 \end{eqnarray}

 \noindent where
 \begin{eqnarray}
 \Sigma _{\theta ,\phi } =  \; i\; \gamma ^{1}
\partial _{\theta } + \gamma ^{2} {i\partial +i\sigma^{12}\cos
\theta  \over \sin \theta  } \; . \nonumber
\end{eqnarray}

\noindent  Eq. (\ref{10.1a}) reads
\begin{eqnarray}
\left [ i\; {\gamma ^{0}  \over \sqrt{\Phi} }   \partial _{t} \; + i
\sqrt{\Phi } \gamma^{3} (
\partial _{r}  +   { 1 \over  r}   +
  { \Phi' \over 4 \Phi  }  \;  )
+   { 1 \over r} \;\Sigma _{\theta,\phi } -  M
  \right ]  \Psi (x)  =  0  \; .
\label{10.1b}
\end{eqnarray}

\noindent With the
substitution
$
 \Psi  (x)  = r ^{-1} \Phi ^{-1/4} \; \psi (x)$,
 we get
\begin{eqnarray}
\left ( i\; {\gamma ^{0}  \over \sqrt{\Phi} }   \partial _{t}  + i
\sqrt{\Phi } \gamma^{3}
\partial _{r}  +   { 1 \over r} \;\Sigma _{\theta,\phi }  -  M
  \right )   \psi (x)  =  0  \; .
\label{10.1c}
\end{eqnarray}

\noindent Below the spinor basis will be used
\begin{eqnarray}
\gamma^{0} = \left | \begin{array}{cc}
0 & I \\
I & 0
\end{array} \right |, \;\;  \gamma^{j} =
\left | \begin{array}{cc}
0 & -\sigma_{j} \\
\sigma_{j}  & 0
\end{array} \right |, \; \;  i \sigma^{12} =
\left | \begin{array}{cc}
\sigma_{3} & 0 \\
0 & \sigma_{3}
\end{array} \right | .
\nonumber
\end{eqnarray}

\section{   Separation of the variables
}

Let us introduce a Dirac string potential in the de Sitter space-time model.
It is convenient to start with  the
monopole Abelian potential in the Schwinger's form for  the flat Minkowski space \cite{Strazhev-Tomilchik}
\begin{eqnarray}
A^{a}(x) = (A^{0}, \; A^{i}) = \left ( 0 \; , \; g\;
 {(\vec{r} \times \vec{n})\;(\vec{r} \; \vec{n}) \over
r \; (r^{2} - (\vec{r} \; \vec{n})^{2}) } \right )  .
\label{3.1a}
\end{eqnarray}

\noindent Specifying $\vec{n} = (0, 0 , 1 )$  and  translating  the $A_{\alpha }(x)$  to  the~spherical
coordinates,  we get
\begin{eqnarray}
A_{0} = 0 , \;\; A_{r} = 0 , \;\; A_{\theta } = 0\; , \qquad
A_{\phi } = g\; \cos \theta  \; .
\label{3.1b}
\end{eqnarray}

\noindent This potential $A_{\phi}$ obeys
Maxwell equations in de Sitter space
\begin{eqnarray}
{1 \over \sqrt{-g}} {\partial \over \partial x^{\alpha}} \sqrt{-g}  F^{\alpha \beta} = 0\; ,
\qquad \sqrt{ -g } = r^{2} \sin \theta\;,
\nonumber
\\
F_{\phi \theta} = g \sin \theta \; ,\;
\;
{1 \over r^{2} \sin \theta }  {\partial \over \partial \theta}  r^{2} \sin \theta\;
{1 \over r^{2}}  {1 \over  r^{2}\sin^{2} \theta} \;  g \sin \theta =0 \; .
\label{3.1c}
\end{eqnarray}

\noindent
Correspondingly,  the~Dirac  equation  in  presence of  this    field  $A_{\phi}$ takes the form
\begin{eqnarray}
\left ( i\; {\gamma ^{0}  \over \sqrt{\Phi} }   \partial _{t} \; + i
\sqrt{\Phi } \gamma^{3}\;   \;
\partial _{r}  +   { 1 \over r} \;\Sigma^{k} _{\theta,\phi }  -  M \right  ) \psi (x)  =  0  \; ,
\label{3.2a}
\end{eqnarray}

\noindent where (below the notation $eg/\hbar c=k$ will be used)
\begin{eqnarray}
\Sigma ^{k}_{\theta ,\phi }  =  i \gamma ^{1} \partial _{\theta}   +
\gamma ^{2}   { i \partial _{\phi } + (i\sigma ^{12} - k )
 \cos \theta \over  \sin  \theta} \; .
\label{3.2b}
\end{eqnarray}

As readily verified, the~wave  operator
  in  (\ref{3.2a}) commutes with the~following three ones
\begin{eqnarray}
J^{k}_{1} =  l_{1} + {(i\sigma ^{12} - k)
 \cos \phi  \over \sin \theta } \; ,
 \nonumber
 \\
J^{k}_{2} = \ l_{2} + {(i\sigma ^{12} - k)
\sin \phi  \over \sin \theta } \; , \;\;\;
\qquad  J^{k}_{3} = l_{3}
\label{3.3a}
\end{eqnarray}

\noindent which  in  turn    obey   the~$su(2)$   Lie   algebra.   Clearly,
this    monopole     situation     comes  entirely  under
the~Schr\"{o}dinger
\cite{Schrodinger-1938}, and Pauli
\cite{Pauli-1939} approach; detailed treatment of the method was given recently in \cite{Book-2};
similar technique
when treating the problem of any spin particle
in magnetic pole  was used  previously   in \cite{Fushchich-Nikitin-Susloparow}, though
with no connection with tetrad formalism.

Corresponding to diagonalization of the
 $\vec{J}^{2}_{k}$ and $J^{k}_{3}$,
the~function $\psi$ is to be taken as
($D_{\sigma } \equiv D^{j}_{-m,\sigma }(\phi ,\theta ,0)$ stands for Wigner functions
\cite{Varshalovich-Moskalev-Hersonskiy-1975})
\begin{eqnarray}
\psi^{k}_{\epsilon jm} (t,r,\theta ,\phi ) = {e^{-i\epsilon t} \over  r} \;
\left |\begin{array}{r}
       f_{1} \; D_{k-1/2}   \\   f_{2} \; D_{k+1/2}   \\
       f_{3} \; D_{k-1/2}   \\   f_{4} \; D_{k+1/2}
\end{array} \right |\; .
\label{3.3b}
\end{eqnarray}

\noindent Further, with the use of  recursive relations  \cite{Varshalovich-Moskalev-Hersonskiy-1975}
\begin{eqnarray}
\partial_{\theta}   D_{k+1/2} = a  D_{k-1/2} - b  D_{k+3/2}
\; ,
\partial_{\theta}   D_{k-1/2} =  c  D_{k-3/2} - a  D_{k+1/2} \; ,
\nonumber
\\
\sin^{-1} \theta   \;[\; -m -(k+1/2) \cos \theta \;] \; D_{k+1/2} =
- a  D_{k-1/2} - b  D_{k+3/2}   \;\; ,
\nonumber
\\
\sin^{-1} \theta  \;[\; -m -(k-1/2)\cos \theta \;] \; D_{k-1/2} =
- c  D_{k-3/2} - a  D_{k+1/2}\; ,
\nonumber
\end{eqnarray}

\noindent where
\begin{eqnarray}
a = {1 \over 2} \sqrt{(j + 1/2)^{2} - k^{2}} \;,
\nonumber
\\
b = { \sqrt{(j - k - 1/2)(j + k + 3/2)}  \over 2} \; ,
\nonumber
\\
c = {\sqrt{(j + k - 1/2)(j - k + 3/2)}  \over 2}  \; ,
\nonumber
\end{eqnarray}

\noindent
we find how the $\Sigma ^{k}_{\theta ,\phi }$  acts on $\psi ^{k}_{\epsilon jm}  $
\begin{eqnarray}
\Sigma ^{k}_{\theta ,\phi } \; \psi ^{k}_{\epsilon jm} =
 i \; \sqrt{(j + 1/2)^{2} - k^{2}} \;\; e^{-i\epsilon t}
\left | \begin{array}{r}
  - f_{4} \; D_{k-1/2}  \\  + f_{3} \; D_{k+1/2} \\
  + f_{2} \; D_{k-1/2}  \\  - f_{1} \; D_{k+1/2}
\end{array} \right |;
\label{3.4}
\end{eqnarray}

\noindent
hereafter the factor $\sqrt{(j + 1/2)^{2}- k^{2}}$
 will be referred to as  $\nu $.
 For the  $f_{i}(r)$, the radial system derived  is
 \begin{eqnarray}
{\epsilon  \over \sqrt{\Phi}} \; f_{3} \; - \; i \; \sqrt{\Phi} {d\over dr} \; f_{3}  \;- \;i\; {\nu \over r}\; f_{4}\;
 - \; M \; f_{1} = 0 \; ,
\nonumber
\\
{\epsilon  \over \sqrt{\Phi}} \; f_{4} \; + \; i\; \sqrt{\Phi}  {d \over dr}\;  f_{4}\; + \;
 i \;{\nu \over r} \; f_{3} \; - \; M \; f_{2} = 0 \; ,
\nonumber
\\
{ \epsilon \over \sqrt{\Phi}} \; f_{1} \; +\; i\; \sqrt{\Phi}{d \over dr} \; f_{1} \;  + \;
i \; {\nu \over r} \; f_{2} \;- \;M\; f_{3} = 0 \; ,
\nonumber
\\
{ \epsilon \over \sqrt{\Phi}} \; f_{2} \; - \;i\; \sqrt{\Phi} {d \over dr}\; f_{2}\; - \;
i \;{\nu \over r}\; f_{1}\; -\; M \;f_{4} = 0 \; .
\label{3.5}
\end{eqnarray}

Else  one  operator  can  be
diagonalized together with $i \partial _{t} , \vec{J}^{2}_{k},
J^{k}_{3} $:  namely, a~generalized Dirac operator
\begin{eqnarray}
\hat{K} ^{k} \; = - \; i \; \gamma^{0} \; \gamma ^{3} \;
\Sigma ^{k}_{\theta ,\phi }   \; .
\label{3.6a}
\end{eqnarray}

\noindent From the  eigenvalue equation $\hat{K}^{k} \psi _{\epsilon jm}  =
\lambda  \; \psi _{\epsilon jm}$    we  can  produce  two
possible values for this $\lambda$  and the corresponding restrictions on $f_{i}(r)$
\begin{eqnarray}
\lambda  = - \delta \;  \sqrt{(j + 1/2)^{2}- k^{2}} \; , \qquad
f_{4} = \delta \;  f_{1} \; , \;\;\; f_{3} = \delta \; f_{2} \; .
\label{3.6b}
\end{eqnarray}

\noindent Correspondingly,  the~system  (\ref{3.5})  reduces to
\begin{eqnarray}
\left ( \sqrt{\Phi} {d \over dr} + {\nu \over r} \right ) f  +
\left ({\epsilon  \over \sqrt{\Phi}} + \delta\;  M  \right )  g = 0\; ,
\nonumber
\\
\left ( \sqrt{\Phi} {d \over dr} - {\nu \over r} \right ) g  -
\left  ({\epsilon  \over \sqrt{\Phi}}  - \delta\;  M  \right )  f = 0\;,
\label{3.7}
\end{eqnarray}

\noindent to exclude  imaginary $i$ we have translated equations to new functions
\begin{eqnarray}
f \; = \; {f_{1} + f_{2} \over \sqrt{2}} \; , \qquad g \; = \;
{f_{1} - f_{2} \over i \sqrt{2}} \; .
\nonumber
\end{eqnarray}

Note the  quantization  rule for   $k = eg/ \hbar c$  and $j$
\begin{eqnarray}
{eg \over hc} = \pm 1/2 , \; \pm 1, \; \pm 3/2, \ldots ;
\nonumber
\\
j = \mid k \mid  -1/2, \mid k \mid +1/2, \mid k \mid +3/2,\ldots
\label{3.8}
\end{eqnarray}

The case of minimal  value $j_{\min}= \mid k \mid - 1/2$  must  be
separated  and treated  in a special way.  For example, let
$k = +1/2$, then to the minimal value $j = 0$ there corresponds a  wave
function in terms of only $(t,r)$-dependent quantities
\begin{eqnarray}
\psi ^{(j=0)}_{k = +1/2}(x) =  e^{-i\epsilon t}
\left | \begin{array}{c}
           f_{1}(r)  \\   0  \\  f_{3}(r)  \\  0
\end{array} \right |   \; ;
\label{3.9a}
\end{eqnarray}

\noindent at $k = - 1/2$,  we have
\begin{eqnarray}
\psi ^{(j =0)}_{k = -1/2}(x) =
e^{-i\epsilon t}
\left | \begin{array}{c}
   0  \\  f_{2}(r)  \\   0   \\  f_{4}(r)
\end{array} \right |       \;          .
\label{3.9b}
\end{eqnarray}

\noindent Thus, if $k = \pm  1/2$, then to the minimal  values
 $j_{\min }$
there correspond the substitutions which do not depend at
all on the angular variables $(\theta ,\phi )$. At
 this point there exists some
formal analogy between  the  electron-monopole  states  and
$S$-states (with $l = 0 $) for a~boson field of spin zero:
$\Phi _{l=0} = \Phi (r,t)$. However, it would be unwise to attach too much
significance
to this formal coincidence  because that $(\theta ,\phi )$-independence
of $(e-g)$-states  is  not the~fact  invariant   under   tetrad   gauge
transformations. In contrast, the relation below (let $k = +1/2)$
\begin{eqnarray}
\Sigma^{+1/2}_{\theta ,\phi }  \psi ^{(j=0)}_{k=+1/2} (x)  =
\gamma ^{2} \;  \cot \theta \; ( i \sigma ^{12} - 1/2 ) \;
\psi ^{(j =0)} _{k=+1/2} \equiv  0
\label{3.10a}
\end{eqnarray}

\noindent is invariant under any gauge transformations. The identity (\ref{3.10a})
holds because all the zeros in the
$\psi ^{(j=0)}_{k=+1/2}$ are adjusted to the non-zeros in
$( i \sigma ^{12} - 1/2 )$  and
conversely;  the  non-vanishing  constituents  in $\psi ^{(j=0)}_{k=+1/2}$
are canceled out by zeros in $( i \sigma ^{12}- 1/2 )$.
Correspondingly, the~matter equation (\ref{3.2a}) assumes  the more simple form
\begin{eqnarray}
\left (   i \; {\gamma ^{0} \over \sqrt{\Phi} }  \partial_{t}  +
 i\; \gamma ^{3} \sqrt{\Phi}  \; \partial_{r}
 -   M \; \right )  \psi ^{(j=0)} = 0\; .
\label{3.10b}
\end{eqnarray}

It is readily  verified  that  both (\ref{3.9a})  and (\ref{3.9b}) representations are
 extended to $(e-g)$-states  with $j = j_{\min }$ at all  other
$k =\pm 1, \pm 3/2, \ldots $ Indeed,
\begin{eqnarray}
k= +1, +3/2, +2,\ldots, \qquad
\psi ^{k > 0} _{j_{\min.}} (x) = e^{-i\epsilon t}
\left | \begin{array}{l}
   f_{1}(r) \; D_{k-1/2}  \\  0  \\  f_{3}(r) \;  D_{k-1/2} \\  0
\end{array} \right | \; ;
\label{4.11a}
\\
k = -1, -3/2,-2,\ldots,  \qquad
\psi ^{k<0} _{j_{\min}} (x) =  e^{-i\epsilon t}
\left | \begin{array}{l}
    0    \\   f_{2}(r) \; D_{k+1/2}  \\  0  \\ f_{4}(r) \; D_{k+1/2}
\end{array} \right | \; ,
\label{4.11b}
\end{eqnarray}

\noindent and  the relation
$\Sigma _{\theta ,\phi } \Psi _{j_{\min }} = 0 $ still  holds.
For instance, let us consider in more detail  the case of positive $k$.
Using the recursive relations
\begin{eqnarray}
\partial _{\theta } D_{k-1/2} =
 { 1 \over 2} \sqrt{ 2k-1}  \; D_{k-3/2}\; ,
 \nonumber
 \\
\sin^{-1} \theta \; [ \;  - m - (k-1/2) \cos \theta \; ] \;   D_{k - 1/2}  =
 - { 1 \over 2} \sqrt{ 2k -1} \; D_{k-3/2} \; ,
\nonumber
\end{eqnarray}

\noindent we get
\begin{eqnarray}
i\gamma ^{1} \; \partial _{\theta}
\left |  \begin{array}{c}
       f_{1}  D_{k-1/2} \\  0  \\  f_{3}  D_{k-1/2}  \\  0
\end{array} \right | = {i\over 2} \sqrt{2k-1} \;
\left | \begin{array}{c}
     0  \\ - f_{3}   D_{k-3/2}  \\ 0  \\ + f_{1} D_{k-3/2}
\end{array} \right | ,
\nonumber
\\
\gamma ^{2}  \; {i\partial _{\phi } + (i\sigma ^{12} - k) \cos \theta \over
\sin \theta} \;
\left | \begin{array}{c}
      f_{1} D_{k-1/2} \\  0  \\  f_{3}  D_{k-1/2} \\ 0
\end{array} \right | \; = \;
{i \over 2} \sqrt{2k-1} \;
\left | \begin{array}{c}
    0 \\ +f_{3} D_{k-3/2}  \\ 0  \\ -f_{1} D_{k-3/2}
\end{array} \right | ;
\nonumber
\end{eqnarray}

\noindent in a~sequence, the identity
$\Sigma _{\theta ,\phi } \; \psi _{j_{\min }} \equiv  0$  takes place.
The  case of negative $k$ can be considered in the same way.
As regards the operator $\hat{K}^{k}$, for the $j_{\min }$  states  we
get
$\hat{K}^{k} \; \psi _{j_{\min}} =  0$.

Thus, at every $k$, the $j_{\min }$-equation  has  the  same
 form
\begin{eqnarray}
\left (  i\; {\gamma ^{0} \over  \sqrt{\Phi}}  \; \partial_{t}  +  i\gamma ^{3} \sqrt{\Phi}\;
\partial_{r}  +  {1 \over r} \;  )   -   M
\right ) \psi _{j_{\min}} = 0 \; ;
\label{3.11c}
\end{eqnarray}

\noindent which leads to the same  radial system

\vspace{3mm}

$
k = +1/2,+1,\ldots $
\begin{eqnarray}
{\epsilon \over \sqrt{\Phi}} \; f_{3} - i \; \sqrt{\Phi} { d\over dr}  \; f_{3}  - M \; f_{1} = 0\; ,
\nonumber
\\
{ \epsilon \over \sqrt{\Phi}} \; f_{1} + i \; \sqrt{\Phi} { d \over dr} \; f_{1}  - M \; f_{3} = 0 \; ;
\label{3.12a}
\end{eqnarray}

$
k = -1/2,-1,\ldots $
\begin{eqnarray}
{\epsilon  \over \sqrt{\Phi}} \; f_{4} + i \; \sqrt{\Phi} { d\over dr}\; f_{4} - M \;f_{2} = 0 \; ,
\nonumber
\\
{ \epsilon \over \sqrt{\Phi}}\; f_{2} - i \; \sqrt{\Phi} { d\over dr}\; f_{2} - M \;f_{4} = 0 \; .
\label{3.12b}
\end{eqnarray}

\noindent
 In the limit of the flat space--time, these equations are equivalent respectively to

$
k = + 1/2,+ 1,\ldots$
\begin{eqnarray}
\left ( {d^{2} \over dr^{2}}  + \epsilon ^{2}  - m^{2}\right ) f_{1}  = 0 \;
, \;\; f_{3} =  { 1 \over m}\left ( \epsilon  +
i { d \over dr }\right )  f_{1}     \; ;
\label{3.13a}
\end{eqnarray}

$ k = - 1/2, - 1,\ldots$
\begin{eqnarray}
\left (
{d^{2} \over dr^{2}}  + \epsilon ^{2}  - m^{2}\right )  f_{4} = 0\;
 ,\;\; f_{2} = {1 \over m} \left ( \epsilon  + i {d \over dr} \right )
  f_{4}\; .
\label{3.13b}
\end{eqnarray}

\noindent These equations   end up  with the functions
 $f = \exp  ( \pm  \sqrt{m^{2} - \epsilon ^{2}} \; r )$.
In particular,  at $\epsilon \; < \; m$, there arise solutions of the form
\begin{eqnarray}
\exp  \; (  -\sqrt{m^{2} -\epsilon ^{2}} \; r \; ) \; ,
\label{4.13c}
\end{eqnarray}

\noindent which seem  to  be appropriate to describe   bound   state s  in
the~electron-monopole system. It  should  be  emphasized  that
today the $j_{\min }$   bound  state  problem   remains   still  yet
a~question to understand.  In  particular,  the  important  question
is of finding a~physical and mathematical   criterion  on
selecting values for $\epsilon $: whether $\epsilon \; < \; m$, or
$\epsilon  = m$ , or $\epsilon \; > \; m$;
and which  value of $\epsilon $  is to be chosen  after specifying an~interval
above. The case $\epsilon =m$ is the most special one  -- it gives $f_{1}=f_{2}= 1$ and
$f_{4}= f_{2}=1$.

\section{  Solution of the radial equations}

 Let us turn back to eqs.
(\ref{3.7}); for definiteness we will consider  the case   $\delta =
+1$ (the case $\delta =-1$ follows from the former through  the
 formal change  $M \Longrightarrow - M$)
\begin{eqnarray}
\left ( \sqrt{\Phi} {d \over dr} + {\nu \over r} \right )  f  +  \left ( {
\epsilon  \over \sqrt{\Phi}}  +   M \right ) g  = 0 \; ,
 \nonumber
\\
\left ( \sqrt{\Phi} {d \over dr}  - {\nu \over r} \right ) g  - \left  (
{\epsilon  \over \sqrt{\Phi}}  -
  M  \right )  f = 0     \; .
\label{10.12}
\end{eqnarray}

 \noindent
 Here  we note  additional singularities at the points
$$
 \epsilon +
\sqrt{\Phi} \;  M  =0   \qquad  \mbox{or} \qquad  \epsilon - \sqrt{\Phi} \;   M
= 0 \;.
$$
For instance, the equation for $f(r)$ has  the form
\begin{eqnarray}
{d^{2}  \over dr^{2}} f - \left ( {2r \over 1 -r^{2} } - {M r
\over  \sqrt{1-r^{2}} (\epsilon + M \sqrt{1 -r^{2}})} \right ) {d
\over dr } f + \left ( { \epsilon^{2} \over (1-r^{2})^{2} } -
{M^{2} \over 1 -r^{2}} \right. \nonumber
\\
\left. -{ \nu (\nu +1) \over r^{2} (1 - r^{2} ) } - {\nu \over
(1-r^{2})   \sqrt{1-r^{2} } } +
 { M \nu  \over  \sqrt{1-r^{2}}   (\epsilon + M \sqrt{1 -r^{2}}) }
\right ) f = 0 \; . \nonumber
\end{eqnarray}

\noindent
However, there exists possibility to move these singularities away
through a special transformation of the functions $f(r), g(r)$ (see in
\cite{Otchik-1985}).
To this end, first let us
introduce a new variable $ r = \sin \rho  $, eqs. (\ref{10.12})
look simpler
\begin{eqnarray}
( {d \over d \rho} \;+\; {\nu \over \sin \rho}\;) \; f \; + \; ( {
\epsilon  \over \cos \rho }  \;+ \;  M )\; g \; = \;0 \; ,
 \nonumber
\\
(  {d \over d \rho} \; - \;{\nu \over \sin \rho}\;)\; g  \;- \; (
{\epsilon  \over \cos \rho }\; - \;
 M )\; f\; =\; 0     \; .
\label{10.13}
\end{eqnarray}

\noindent Summing and subtracting two  last equations, we get
\begin{eqnarray}
{d \over d \rho} (f+g) + {\nu \over \sin \rho} (f-g) -  {\epsilon
\over \cos \rho} (f-g) + M (f+g) = 0 \; , \nonumber
\\
{d \over d \rho} (f-g) + {\nu \over \sin \rho} (f+g) +  {\epsilon
\over \cos \rho} (f+g) - M(f-g) = 0 \; . \label{10.14}
\end{eqnarray}

\noindent Introducing two  new  functions
\begin{eqnarray}
f + g = e^{-i\rho/2} (F + G) \; , \qquad f - g = e^{+i\rho/2} (F -
G) \; , \label{10.15}
\end{eqnarray}

\noindent or in matrix form
\begin{eqnarray}
\left | \begin{array}{c}
f \\
g
\end{array} \right | =
\left | \begin{array}{cc}
\cos {\rho \over 2} & - i \sin {\rho \over 2} \\
- i \sin {\rho \over 2} & \cos {\rho \over 2}
\end{array} \right |
\left | \begin{array}{c}
F \\
G
\end{array} \right |
\end{eqnarray}

\noindent one transforms (\ref{10.14})  into
\begin{eqnarray}
{d \over d \rho} e^{-i\rho/2} (F + G)  + {\nu \over \sin \rho}
e^{+i\rho/2} (F - G) \nonumber
\\
-  {\epsilon \over \cos \rho}  e^{+i\rho /2} (F - G)  + M e^{-i
\rho /2} (F + G)  = 0 \; , \nonumber
\\
{d \over d \rho} e^{+i\rho /2} (F - G)  + {\nu \over \sin \rho}
e^{-i\rho /2} (F + G) \nonumber
\\
+  {\epsilon \over \cos \rho} e^{-i \rho /2} (F + G)  - M e^{+i
\rho /2} (F - G)  = 0 \; , \nonumber
\end{eqnarray}

\noindent or
\begin{eqnarray}
 {d \over d \rho}  (F + G)  - {i \over 2} (F + G) +
{\nu \over \sin \rho} (\cos \rho + i \sin \rho)  (F - G)
 \nonumber
\\
-  {\epsilon \over \cos \rho}  (\cos \rho + i \sin \rho)  (F - G)
+ M  (F + G)  = 0 \; , \nonumber
\end{eqnarray}
\begin{eqnarray}
{d \over d \rho}  (F - G)  + {i \over 2}  (F - G)  + {\nu \over
\sin \rho}  (\cos \rho - i \sin \rho) (F + G)
 \nonumber
\\
+  {\epsilon \over \cos \rho} (\cos \rho - i \sin \rho) (F + G)  -
M  (F - G)  = 0 \; . \nonumber
\end{eqnarray}

\noindent Finally, summing and subtracting two last  equations
\begin{eqnarray}
 ({d  \over d \rho}    + \nu \;  { \cos \rho \over \sin \rho}   - i \epsilon \; { \sin \rho \over  \cos  \rho}
 ) \; F
+ \;  ( \; \epsilon  + M  - i \nu   - {i \over 2}   )\;  G = 0 \;
, \nonumber
\\
({d  \over d \rho}    - \nu   \; { \cos \rho \over \sin \rho}
 + i  \epsilon \; { \sin \rho \over  \cos  \rho}  ) \; G
+ ( - \epsilon  + M   + i \nu   - {i \over 2}  )  \; F = 0 \; .
\label{10.17}
\end{eqnarray}

\noindent
The equations produced have no  singular points, except $\rho =0 , \pi /2$.
It should be noted that for the  second class of states with $\delta=-1$,
the relevant system differ from (\ref{10.17}) in  the  sign of $M$:  $M \Longrightarrow -M$.

Let us translate the system  (\ref{10.17}) to the variable
\begin{eqnarray}
z = r^{2} = \sin^{2} \rho\;  , \qquad z \in [0, +1 ) \; ;
\label{variable}
\end{eqnarray}
\begin{eqnarray}
\left | \begin{array}{c}
f \\
g
\end{array} \right | =
\left | \begin{array}{cc}
\sqrt{{1 + \sqrt{1-z}\over 2}}
& - i \sqrt{{1 - \sqrt{1-z}\over 2}} \\
- i \sqrt{{1 - \sqrt{1-z}\over 2}} & \sqrt{{1 + \sqrt{1-z}\over 2}}
\end{array} \right |
\left | \begin{array}{c}
F \\
G
\end{array} \right |,
\end{eqnarray}

\noindent
so we get
\begin{eqnarray}
 ( 2\sqrt{z(1-z)} {d  \over d z}    + \nu   { \sqrt{1-z} \over \sqrt{z} }   - i \epsilon \; { \sqrt{z} \over
 \sqrt{1-z}  }
 )  F
\nonumber
\\
 + ( + \epsilon  + M  - i \nu   - {i \over 2}   )\;  G = 0 \;
, \nonumber
\\
(2\sqrt{z(1-z)} {d  \over d z}   - \nu    { \sqrt{1-z}  \over \sqrt{z}}
 + i  \epsilon  { \sqrt{z}  \over  \sqrt{1-z} }  )  G
\nonumber
\\
+ ( - \epsilon  + M   + i \nu   - {i \over 2}  )   F = 0 \; .
\label{10.17}
\end{eqnarray}

From  (\ref{10.17}) it follow  2-nd order differential equations
for  $F$ and  $G$
\begin{eqnarray}
z(1-z){d^{2}F\over dz^{2}}+ ({1\over 2}-z )
{dF\over dz}
\nonumber
\\
+\left[-{1\over 4}\left(M-{i\over 2}\right)^{2}+{\epsilon(\epsilon-i)\over 4(1-z)}-
{\nu(\nu+1)\over 4z}\right]F=0\,,
\nonumber \\
z(1-z){d^{2}G\over dz^{2}}+ ({1\over 2}-z ){dG\over dz}
\nonumber
\\
+
\left[-{1\over 4}\left(M-{i\over 2}\right)^{2}+
{\epsilon(\epsilon+i)\over 4(1-z)}-{\nu(\nu-1)\over 4z}\right]G=0\,.
\label{M10.17}
\end{eqnarray}

\noindent
It should be noted  symmetry between two equations according to formal  changes
\begin{eqnarray}
\nu \longrightarrow - \nu\;, \qquad \epsilon \longrightarrow - \epsilon \; .
\label{M10.17'}
\end{eqnarray}

Let us introduce substitutions
\begin{eqnarray}
F=z^{A}(1-z)^{B}\bar{F}(z)\,, \qquad
G=z^{K}(1-z)^{L}\bar{G}(z)\,,
\nonumber
\end{eqnarray}

\noindent eqs.  (\ref{M10.17}) give
\begin{eqnarray}
z(1-z)\,{d^{2} \bar{F}\over dz^{2}}+\left[2A+{1\over 2}-(2A+2B+1)z\right]\,{d\bar{F}\over dz}
\nonumber
\\
+\left[-{1\over 4}\left(M-{i\over 2}\right)^{2}-
(A+B)^{2}+{\epsilon(\epsilon-i)+2B(2B-1)\over 4(1-z)} \right.
\nonumber
\\
\left.
- {\nu(\nu+1)-2A(2A-1)\over 4z}\right] \bar{F}=0\,,
\label{M1}
\end{eqnarray}

\vspace{3mm}
\begin{eqnarray}
z(1-z)\,{d^{2}\bar{G}\over dz^{2}}+\left[2K+{1\over 2}-(2K+2L+1)z\right]\,{d\bar{G}\over dz}
\nonumber
\\
+\left[-{1\over 4}\left(M-{i\over 2}\right)^{2}-
(K+L)^{2}+{\epsilon(\epsilon+i)+2L(2L-1)\over 4(1-z)} \right.
\nonumber
\\
\left. -
{\nu(\nu-1)-2K(2K-1)\over 4z}\right]\bar{G}=0\,.
\label{M2}
\end{eqnarray}

First let us consider eq.  (\ref{M1}); at $A,B$ taken accordingly
\begin{eqnarray}
A={1+\nu\over 2}\,,\;-{\nu\over 2}\,,\qquad B=-{i\epsilon\over 2}\,,\;{1+i\epsilon\over 2}
\label{M3a}
\end{eqnarray}

\noindent it becomes simpler
\begin{eqnarray}
z(1-z)\,{d^{2} \bar{F}\over dz^{2}}+\left[2A+{1\over 2}-(2A+2B+1)z\right]\,{d\bar{F}\over dz}
\nonumber
\\
+\left[-{1\over 4}\left(M-{i\over 2}\right)^{2}-
(A+B)^{2}\right] \bar{F}=0\,,
\label{M3b}
\end{eqnarray}

\noindent which is of hypergeometric type with parameters
$$
a= A+B + {iM +1/2 \over 2}\,,\qquad b= A+B -{iM+1/2\over 2} \,, \qquad  c = 2A +1/2 \; .
$$

To have  solutions regular in the origin $z=0$, we should  take positive $A$. Thus there arise two sorts of solutions
depending on a chosen B (in each case two linearly independent solutions, regular and singular in the origin,
 are written down):

\vspace{3mm}
the first
\begin{eqnarray}
A+B ={1+\nu -i \epsilon \over 2} \, \qquad c = \nu +3/2\; ,
\nonumber
\\
\bar{F}^{(1)}_{reg} (z) = F(a,b,c;z) \;,
\nonumber
\\
\bar{F}^{(1)}_{sing} (z) = z^{1-c} F(a+1-c,b+1-c,2-c;z) \;,
\nonumber
\\
a = {1+\nu -i \epsilon \over 2} + {iM +1/2 \over 2} \;,
\nonumber
\\
b = {1+\nu -i \epsilon \over 2} - {iM +1/2 \over 2 }\; ;
\label{M3c}
\end{eqnarray}

\vspace{3mm}
the second
\begin{eqnarray}
A+B ={2+\nu +i \epsilon \over 2} \, \qquad \gamma  = \nu +3/2\; ,
\nonumber
\\
\bar{F}^{(2)}_{reg} (z) = F(\alpha,\beta, \gamma;z) \;,
\nonumber
\\
\bar{F}^{(2)}_{sing} (z) = z^{1-\gamma} F(\alpha +1-\gamma, \beta +1-\gamma,2- \gamma;z) \;,
\nonumber
\\
\alpha  = {2+\nu +i \epsilon \over 2} + {iM +1/2 \over 2}\; ,
\nonumber
\\
\beta = {2+\nu +i \epsilon \over 2} - {iM +1/2 \over 2 }\; .
\label{M3c}
\end{eqnarray}

Not let us turn back to eq. (\ref{M2}); at  $K$ and $L$ chosen according to
\begin{eqnarray}
K={1-\nu\over 2}\,,\;{\nu\over 2}\,,\qquad L={i\epsilon\over 2}\,,\;{1-i\epsilon\over 2}
\label{M4a}
\end{eqnarray}

\noindent it will be simpler
\begin{eqnarray}
z(1-z)\,{d^{2}\bar{G}\over dz^{2}}+\left[2K+{1\over 2}-(2K+2L+1)z\right]\,{d\bar{G}\over dz}
\nonumber
\\
+\left[-{1\over 4}\left(M-{i\over 2}\right)^{2}-
(K+L)^{2}\right]\bar{G}=0\,,
\label{M4b}
\end{eqnarray}

\noindent which is of hypergeometric type
\begin{eqnarray}
a'=K+L +{iM+1/2\over 2} \,,\qquad b'= K+L  -{iM+1/2\over 2}\,, \qquad c' = 2K +{1\over 2} \; .
\nonumber
\end{eqnarray}

To have  solutions regular in the origin $z=0$, we take positive $K$. Thus there arise two sorts of solutions
depending on a chosen B (in each case two linearly independent solutions, regular and singular in the origin,
 are written down)

\vspace{2mm}
the first
\begin{eqnarray}
K+L ={\nu +i \epsilon \over 2} \, \qquad c' = \nu +1/2  \; ,
\nonumber
\\
\bar{G}^{(1)}_{reg} (z) = F(a',b',c';z) \;,
\nonumber
\\
\bar{G}^{(1)}_{sing} (z) = z^{1-c'} F(a'+1-c',b'+1-c',2-c';z) \;,
\nonumber
\\
a' = {\nu +i \epsilon \over 2} + {iM +1/2 \over 2} \; ,
\nonumber
\\
b' = {\nu +i \epsilon \over 2} - {iM +1/2 \over 2 }  \; ;
\label{M3c}
\end{eqnarray}

the second
\begin{eqnarray}
K+L ={\nu+1  -i \epsilon \over 2} \, \qquad \gamma ' = \nu +1/2
 \; ,
\nonumber
\\
\bar{G}^{(2)}_{reg} (z) = F(\alpha',\beta', \gamma';z)   \;,
\nonumber
\\
\bar{G}^{(2)}_{sing} (z) = z^{1-\gamma'} F(\alpha' +1-\gamma', \beta' +1-\gamma',2- \gamma';z) \;,
\nonumber
\\
\alpha'  = {\nu+1  -i \epsilon \over 2} + {iM +1/2 \over 2}  =a   \; ,
\nonumber
\\
\beta'  = {\nu +1  -i \epsilon \over 2} - {iM +1/2 \over 2 }  = b   \; .
\label{M3c}
\end{eqnarray}

Thus, we have constructed the following four regular solutions
\begin{eqnarray}
F^{(1)}_{reg}   \; , \qquad  F^{(2)}_{reg} , \qquad G^{(1)}_{reg} \;, \qquad G^{(2)}_{reg} \; .
\nonumber
\end{eqnarray}

\noindent
However, due to the known identity for hypergeometric functions
\begin{eqnarray}
F(A,B,C;z) = (1-z)^{C-A-B} F(C-A,C-B,C;z)
\label{*}
\end{eqnarray}

\noindent
we readily conclude that there exist only two different ones:
\begin{eqnarray}
F^{(1)}_{reg}  =z ^{(\nu+1)/2} (1-z)^{-i\epsilon/2} F(a,b,c,z)
\nonumber
\\
 =z ^{(\nu+1)/2} (1-z)^{(1+i\epsilon)/2} F(\alpha ,\beta,\gamma,z) =F^{(2)}_{reg}    \; ,
\end{eqnarray}
\begin{eqnarray}
G^{(1)}_{reg}  =z ^{\nu/2} (1-z)^{+i\epsilon/2} F(a',b',c',z)
\nonumber
\\
  =z ^{\nu/2} (1-z)^{(1-i\epsilon)/2} F(\alpha' ,\beta',\gamma',z)  =G^{(2)}_{reg} \; .
\end{eqnarray}

\noindent
The same is true for singular solutions
\begin{eqnarray}
F^{(1)}_{sing}  =   F^{(2)}_{sing}  \; , \qquad G^{(1)}_{sing} =  G^{(2)}_{sing} \; .
\label{idetity}
\end{eqnarray}

Taking into account relations
\begin{eqnarray}
 \alpha = a' +1     , \qquad
  \beta  = b' +1  \; ,  \qquad   \gamma = c' +1   \; , \nonumber
  \\
 \bar{G}^{(1)}_{reg} (z)   \qquad \Longrightarrow  \qquad  \bar{F}_{reg}^{(2)}  (z)  \; ;
\label{identities}
\end{eqnarray}
we can expect  an identity
\begin{eqnarray}
 \left ( 2\sqrt{z(1-z)} {d  \over d z}    - \nu   { \sqrt{1-z} \over \sqrt{z} }   + i \epsilon \; { \sqrt{z} \over
 \sqrt{1-z}  }\right  )
 \nonumber
 \\
 \times
 G_{0} ^{reg}   z^{\nu/2}(1-z)^{+i\epsilon/2} F(a',b',c',z)
\nonumber
\\
 + ( - \epsilon  + M  + i \nu   - {i \over 2}   )\;
F_{0}^{reg}  z^{(1+\nu)/2} (1-z)^{(1+i\epsilon)/2} F(\alpha, \beta, \gamma,z)
 = 0 \; .
\nonumber
\end{eqnarray}

\noindent
From whence it follows
\begin{eqnarray}
2G_{0}^{reg} {d \over dz} F(a',b',c',z)  +
( - \epsilon  + M  + i \nu   - {i \over 2}   )\;
F_{0}^{reg}  F(\alpha, \beta, \gamma,z) =0\; ,
\end{eqnarray}

\noindent
and further
\begin{eqnarray}
2G_{0}^{reg} {a' b' \over  c'}  +
( - \epsilon  + M  + i \nu   - {i \over 2}   )
F_{0}^{reg}  =0\;  .
\end{eqnarray}

In the same manner, noting that
\begin{eqnarray}
 (\alpha ' + 1 - \gamma ') = (a+1 - c)  +1    \;,
\nonumber
\\
 (\beta ' + 1 - \gamma ') = (b+1 - c) +1     \;,
\nonumber
\\
(2-\gamma' ) = (2-c) +1    \; ,
\nonumber
  \\
\bar{F}^{(1)}_{sing} (z)   \qquad \Longrightarrow  \qquad  \bar{G}_{sing}^{(2)}  (z)    \;,
 \label{identities'}
\end{eqnarray}

\noindent
 we can assume that
\begin{eqnarray}
 \left (2\sqrt{z(1-z)} {d  \over d z}   + \nu    { \sqrt{1-z}  \over \sqrt{z}}
 - i  \epsilon  { \sqrt{z}  \over  \sqrt{1-z} }  \right  )
 \nonumber
 \\
 \times
  F_{0}^{sing} z^{-\nu/2} (1-z)^{-i \epsilon/2}
  F(a+1-c, b+1-c, 2-c,z)
\nonumber
\\
+(  \epsilon  + M   - i \nu   - {i \over 2}  ) z^{(1 - \nu)/2} (1-z)^{(1-i\epsilon)/2}
\nonumber
\\
\times
G_{0}^{sing} F(\alpha' +1-\gamma', \beta' +1-\gamma',2- \gamma';z ) = 0 \; .
\end{eqnarray}

\noindent From this it follows
\begin{eqnarray}
2F_{0}^{sing} {d \over dz} F(a+1-c, b+1-c, 2-c,z)
\nonumber
\\
+ (  \epsilon  + M   - i \nu   - {i \over 2}  )
G_{0} ^{sing} F(\alpha' +1-\gamma', \beta' +1-\gamma',2- \gamma';z ) = 0 \; ,
\end{eqnarray}

\noindent which leads us to
\begin{eqnarray}
2F_{0}^{sing} {(a+1-c)(  b+1-c) \over  2-c} +
(  \epsilon  + M   - i \nu   - {i \over 2}  ) G_{0} ^{sing}=0
\nonumber
\end{eqnarray}

\noindent
so that
\begin{eqnarray}
F_{0} ^{sing} (-i\epsilon - \nu + iM +1/2) + i (1-2\nu) G_{0} ^{sing}= 0 \; .
\end{eqnarray}

Thus, we have constructed regular and singular solutions of the system:
\begin{eqnarray}
F^{(1)}_{reg}   =  F^{(2)}_{reg}= F_{reg} \qquad  -- \qquad   G_{reg} =   G^{(1)}_{reg} = G^{(2)}_{reg}  \; ;
\nonumber\\[3mm]
F^{(1)}_{sing}  =   F^{(2)}_{sing} = F_{sing}  \qquad  -- \qquad     G_{sing} =  G^{(1)}_{sing} =  G^{(2)}_{sing}  \; .
\label{idetity}
\end{eqnarray}

\section{  Radial equations in the case $j_{\min}$ }

Let us turn back to the case of the minimal value of $j$:

\vspace{3mm}

$
k = +1/2,+1,\ldots $
\begin{eqnarray}
{\epsilon \over \sqrt{\Phi}} \; f_{3} - i \; \sqrt{\Phi} { d\over dr}  \; f_{3}  - M \; f_{1} = 0\; ,
\nonumber
\\
{ \epsilon \over \sqrt{\Phi}} \; f_{1} + i \; \sqrt{\Phi} { d \over dr} \; f_{1}  - M \; f_{3} = 0 \; ;
\label{5.1a}
\end{eqnarray}

\noindent from whence for new functions
\begin{eqnarray}
h \; = \; {f_{1} + f_{3} \over \sqrt{2}} \; , \qquad g \; = \;
{f_{1} - f_{3} \over i \sqrt{2}} \;
\label{functions-1}
\end{eqnarray}

\noindent
we derive

$
k = +1/2,+1,\ldots $
\begin{eqnarray}
\sqrt{\Phi} {d \over dr} h + \left ( {\epsilon \over \sqrt{\Phi}} + M \right ) g = 0\; ,
\qquad
\sqrt{\Phi} {d \over dr} g - \left ( {\epsilon \over \sqrt{\Phi}} - M \right ) h = 0 \; .
\label{5.1b}
\end{eqnarray}

In the same manner for another case we have

\vspace{3mm}

$
k = -1/2,-1,\ldots $
\begin{eqnarray}
{\epsilon  \over \sqrt{\Phi}} \; f_{4} + i \; \sqrt{\Phi} { d\over dr}\; f_{4} - M \;f_{2} = 0 \; ,
\nonumber
\\
{ \epsilon \over \sqrt{\Phi}}\; f_{2} - i \; \sqrt{\Phi} { d\over dr}\; f_{2} - M \;f_{4} = 0 \; ;
\label{5.2a}
\end{eqnarray}

\noindent  for new functions (note difference between  (\ref{functions-1})
 and (\ref{functions-2}))

\begin{eqnarray}
g \; = \; {f_{2} + f_{4} \over \sqrt{2}} \; , \qquad h \; = \;
{f_{2} - f_{4} \over i \sqrt{2}} \;
\label{functions-2}
\end{eqnarray}
we obtain
\begin{eqnarray}
\sqrt{\Phi} {d \over dr} h + \left ( {\epsilon \over \sqrt{\Phi}} - M \right ) g = 0\; ,
\qquad
\sqrt{\Phi} {d \over dr} g - \left ( {\epsilon \over \sqrt{\Phi}} + M \right ) h = 0 \; .
\label{5.2b}
\end{eqnarray}

Let us  perform special transformation over  the  functions
\begin{eqnarray}
g + h = e^{-i \rho /2 } (F + G) \;, \qquad
g - h = e^{+i \rho /2 } (F - G).
\label{5.4a}
\end{eqnarray}

\noindent
After simple calculation we arrive  at

\vspace{3mm}
\underline{when $k = +1/2,+3/2, ...$}
\begin{eqnarray}
 \left ({d  \over d \rho}      - i \epsilon \; { \sin \rho \over  \cos  \rho}
\right  )  F
+ \;  \left (  \epsilon  + M    - {i \over 2}  \right  )  G = 0 \;
, \;\nonumber
\\
\left ({d  \over d \rho}
 + i  \epsilon \; { \sin \rho \over  \cos  \rho} \right )  G
+ \left  ( - \epsilon  + M     - {i \over 2} \right )    F = 0 \; ;
\label{5.5a}
\end{eqnarray}

\underline{when $k = -1/2,-3/2, ...$}
\begin{eqnarray}
\left  ({d  \over d \rho}      - i \epsilon \; { \sin \rho \over  \cos  \rho}
 \right )  G
+ \;  \left (  \epsilon  - M    - {i \over 2}  \right  )  H = 0 \;
, \;\nonumber
\\
\left ({d  \over d \rho}
 + i  \epsilon \; { \sin \rho \over  \cos  \rho} \right  )  H
+ \left ( - \epsilon  - M     - {i \over 2}  \right )  G = 0 \; .
\label{5.5b}
\end{eqnarray}

The difference between  (\ref{5.5a})  and (\ref{5.5b}) consists in  the only change $M \longleftrightarrow - M$.
The system (\ref{5.5a}) can be compared with the similar one  (\ref{10.17})
\begin{eqnarray}
\left  ({d  \over d \rho}    + \nu \;  { \cos \rho \over \sin \rho}   - i \epsilon \; { \sin \rho \over  \cos  \rho}
 \right ) F
+ \left (  \epsilon  - i \nu   + M    - {i \over 2}  \right )  G = 0 \;
, \;\; \nonumber
\\
\left ({d  \over d \rho}    - \nu   \; { \cos \rho \over \sin \rho}
 + i  \epsilon \; { \sin \rho \over  \cos  \rho}  \right )  G
+ \left ( - \epsilon  + i \nu  + M      - {i \over 2} \right )   F = 0 \; .
\label{5.5c}
\end{eqnarray}

We immediately conclude that  the approach used to treat  (\ref{5.5c}) can be applied  here as well.
In particular,  the system (\ref{5.5a}) being translated to the  variable $z$
\begin{eqnarray}
\sin \rho = \sqrt{z} \; , \qquad \cos \rho = \sqrt{1 - z } \; ,
\qquad
{d \over d \rho} = 2\sqrt{z(1-z)} {d \over dz} \; ,
\nonumber
\end{eqnarray}

\noindent
 will take the form
\begin{eqnarray}
\sqrt{z(1-z)}  \left ( {d \over dz} - {i\epsilon /2 \over 1-z} \right ) F + {M+\epsilon - i/2 \over 2} \;G = 0 \; ,
\nonumber
\\
\sqrt{z(1-z)}  \left ( {d \over dz} + {i\epsilon /2 \over 1-z} \right ) G + {M-\epsilon - i/2 \over 2} \; F = 0 \; .
\label{5.6}
\end{eqnarray}

Because to states with minimal $j$ are  drawn usually   great attention,
let us specify these states in  more detail.
From  (\ref{5.6}) it follow  2-nd order differential equations
for  $F$ and  $G$ respectively
\begin{eqnarray}
z(1-z){d^{2}F\over dz^{2}}+ ({1\over 2}-z )
{dF\over dz}+\left[-{1\over 4}\left(M-{i\over 2}\right)^{2}+{\epsilon(\epsilon-i)\over 4(1-z)} \right]F=0\,,
\nonumber \\
z(1-z){d^{2}G\over dz^{2}}+ ({1\over 2}-z ){dG\over dz}+
\left[-{1\over 4}\left(M-{i\over 2}\right)^{2}+
{\epsilon(\epsilon+i)\over 4(1-z)} \right]G=0\,.
\nonumber
\\
\label{5.7}
\end{eqnarray}

\noindent
It should be noted  symmetry between two equations according to formal  changes
$
 \epsilon \longrightarrow - \epsilon $.
Let us introduce substitutions
\begin{eqnarray}
F=z^{A}(1-z)^{B}\bar{F}(z)\,, \qquad
G=z^{K}(1-z)^{L}\bar{G}(z)\,,
\nonumber
\end{eqnarray}

\noindent eqs.  (\ref{5.7}) give
\begin{eqnarray}
z(1-z)\,{d^{2} \bar{F}\over dz^{2}}+\left[2A+{1\over 2}-(2A+2B+1)z\right]\,{d\bar{F}\over dz}
\nonumber
\\
+\left[-{1\over 4}\left(M-{i\over 2}\right)^{2}-
(A+B)^{2}+{\epsilon(\epsilon-i)+2B(2B-1)\over 4(1-z)} \right.
\nonumber
\\
\left.
+ {2A(2A-1)\over 4z}\right] \bar{F}=0\,,
\label{5.8}
\end{eqnarray}
\begin{eqnarray}
z(1-z)\,{d^{2}\bar{G}\over dz^{2}}+\left[2K+{1\over 2}-(2K+2L+1)z\right]\,{d\bar{G}\over dz}
\nonumber
\\
+\left[-{1\over 4}\left(M-{i\over 2}\right)^{2}-
(K+L)^{2}+{\epsilon(\epsilon+i)+2L(2L-1)\over 4(1-z)} \right.
\nonumber
\\
\left. + {2K(2K-1)\over 4z}\right]\bar{G}=0\,.
\label{5.9}
\end{eqnarray}

First let us consider eq.  (\ref{5.8}); at $A$ and $B$ taken accordingly
\begin{eqnarray}
A={1\over 2}\,,\;0 \,,\qquad B = -{i\epsilon\over 2}\,,\;{1+i\epsilon\over 2}
\label{5.10}
\end{eqnarray}

\noindent it becomes simpler
\begin{eqnarray}
z(1-z)\,{d^{2} \bar{F}\over dz^{2}}+\left[2A+{1\over 2}-(2A+2B+1)z\right]\,{d\bar{F}\over dz}
\nonumber
\\
+\left[-{1\over 4}\left(M-{i\over 2}\right)^{2}-
(A+B)^{2}\right] \bar{F}=0\,,
\label{5.11}
\end{eqnarray}

\noindent which is of hypergeometric type with parameters
$$
a= A+B + {iM +1/2 \over 2}\,,\qquad b= A+B -{iM+1/2\over 2} \,, \qquad  c = 2A +1/2 \; .
$$

\noindent
To have  solutions non-vanishing  in the origin $z=0$, we take zero $A=0$. Thus there arise two sorts of solutions
depending on a chosen B (in each case two linearly independent solutions, regular and singular in the origin,
 are written down)

\vspace{3mm}
the first
\begin{eqnarray}
A+B ={ -i \epsilon \over 2} \, \qquad c =  +1/2\; ,
\nonumber
\\
\bar{F}^{(1)}_{non-zero} (z) = F(a,b,c;z) \;,
\nonumber
\\
\bar{F}^{(1)}_{zero} (z) = z^{1-c} F(a+1-c,b+1-c,2-c;z) \;,
\nonumber
\\
a = { -i \epsilon \over 2} + {iM +1/2 \over 2} ,
\qquad
b = { -i \epsilon \over 2} - {iM +1/2 \over 2 }\; ;
\label{5.12}
\end{eqnarray}

\vspace{3mm}
the second
\begin{eqnarray}
A+B ={1 +i \epsilon \over 2} \, \qquad \gamma  = +1/2\; ,
\nonumber
\\
\bar{F}^{(2)}_{non-zero} (z) = F(\alpha,\beta, \gamma;z) \;,
\nonumber
\\
\bar{F}^{(2)}_{zero} (z) = z^{1-\gamma} F(\alpha +1-\gamma, \beta +1-\gamma,2- \gamma;z) \;,
\nonumber
\\
\alpha  = {1 +i \epsilon \over 2} + {iM +1/2 \over 2} ,\qquad
\beta = {1 +i \epsilon \over 2} - {iM +1/2 \over 2 }\; ;
\label{5.13}
\end{eqnarray}

Now  let us turn back to eq. (\ref{5.9}); at  $K$ and $L$ chosen according to
\begin{eqnarray}
K={1\over 2}\,,\;0\,,\qquad L={i\epsilon\over 2}\,,\;{1-i\epsilon\over 2}
\label{5.14}
\end{eqnarray}

\noindent it will be simpler
\begin{eqnarray}
z(1-z)\,{d^{2}\bar{G}\over dz^{2}}+\left[2K+{1\over 2}-(2K+2L+1)z\right]\,{d\bar{G}\over dz}
\nonumber
\\
+\left[-{1\over 4}\left(M-{i\over 2}\right)^{2}-
(K+L)^{2}\right]\bar{G}=0\,,
\label{5.15}
\end{eqnarray}

\noindent which is of hypergeometric type
\begin{eqnarray}
a'=K+L +{iM+1/2\over 2} \,,\qquad b'= K+L  -{iM+1/2\over 2}\,, \qquad c' = 2K +{1\over 2} \; .
\nonumber
\end{eqnarray}

We start with   solutions non-vanishing in the origin $z=0$, we take zero  $K=0$. Thus there arise two sorts of solutions
depending on a chosen B (in each case two linearly independent solutions, regular and singular in the origin,
 are written down)

\vspace{3mm}
the first
\begin{eqnarray}
K+L ={ +i \epsilon \over 2} \, \qquad c' =  +1/2  \; ,
\nonumber
\\
\bar{G}^{(1)}_{non-zero} (z) = F(a',b',c';z) \;,
\nonumber
\\
\bar{G}^{(1)}_{zero} (z) = z^{1-c'} F(a'+1-c',b'+1-c',2-c';z) \;,
\nonumber
\\
a' = { +i \epsilon \over 2} + {iM +1/2 \over 2} \; ,
\qquad b' = { +i \epsilon \over 2} - {iM +1/2 \over 2 }  \; ;
\label{5.16}
\end{eqnarray}

\vspace{3mm}
the second
\begin{eqnarray}
K+L ={1  -i \epsilon \over 2} \, \qquad \gamma ' =  +1/2
 \; ,
\nonumber
\\
\bar{G}^{(2)}_{non-zero} (z) = F(\alpha',\beta', \gamma';z)   \;,
\nonumber
\\
\bar{G}^{(2)}_{zero} (z) = z^{1-\gamma'} F(\alpha' +1-\gamma', \beta' +1-\gamma',2- \gamma';z) \;,
\nonumber
\\
\alpha'  = {1  -i \epsilon \over 2} + {iM +1/2 \over 2}     ,\qquad
\beta'  = { 1  -i \epsilon \over 2} - {iM +1/2 \over 2 }     \; ;
\label{5.17}
\end{eqnarray}

Thus, we have constructed the following four regular solutions
\begin{eqnarray}
F^{(1)}_{non-zero}   \; , \qquad  F^{(2)}_{non-zero} , \qquad G^{(1)}_{non-zero} \;, \qquad G^{(2)}_{non-zero} \; ;
\nonumber
\end{eqnarray}

\noindent
Due to the known identity for hypergeometric functions
\begin{eqnarray}
F(A,B,C;z) = (1-z)^{C-A-B} F(C-A,C-B,C;z)
\nonumber
\end{eqnarray}

\noindent
we readily conclude that there exist only two different ones
\begin{eqnarray}
F^{(1)}_{non-zero}  = F^{(2)}_{non-zero}    \; ,\qquad
G^{(1)}_{non-zero}  = G^{(2)}_{non-zero} \; .
\end{eqnarray}

The same is true for zero-solutions
\begin{eqnarray}
F^{(1)}_{zero}  =   F^{(2)}_{zero}  \; , \qquad G^{(1)}_{zero} =  G^{(1)}_{zero} \; .
\label{5.19}
\end{eqnarray}

Due to
\begin{eqnarray}
F^{(1)}_{non-zero} = F_{0}^{non-zero} (1-z)^{-i\epsilon/2} F(a,b,c,z) \; ,
\nonumber
\\
G^{(2)}_{zero} = G_{0}^{zero} z^{1/2} (1-z) ^{(1-i\epsilon)/2} F(\alpha' +1-
\gamma', \beta'+1- \gamma',2- \gamma', z) \; ,
\nonumber
\\
a+1 = \alpha' +1- \gamma' ,\;
b+1 =  \beta'+1- \gamma' ,\;
c+1 = 2- \gamma' \; ,
\label{5.20a}
\end{eqnarray}

\noindent
 we  can assume relationship
\begin{eqnarray}
2\sqrt{z(1-z)}  \left ( {d \over dz} - {i\epsilon /2 \over 1-z} \right ) F^{(1)}_{non-zero}
 + ( M+\epsilon - i/2 )  \;G^{(2)}_{zero} = 0 \; .
\label{5.20b}
\end{eqnarray}

\noindent
Indeed, from (\ref{5.20b})  we readily derive
\begin{eqnarray}
2 {ab \over c} F^{non-zero}_{0} +  ( M+\epsilon - i/2 )  \;G^{zero}_{0}  = 0 \qquad \Longrightarrow
\nonumber
\\
a \;  F^{non-zero}_{0} +  ic   \;G^{zero}_{0}  = 0 \; .
\end{eqnarray}

And similarly, due to
\begin{eqnarray}
G^{(1)}_{non-zero} = G_{0}^{non-zero} (1-z)^{+i\epsilon/2} F(a',b',c',z) \; ,
\nonumber
\\
F^{(2)}_{zero} = F_{0}^{zero}
 z^{1/2} (1-z) ^{(1+i\epsilon)/2} F(\alpha +1- \gamma, \beta+1- \gamma ,2- \gamma, z) \; ,
\nonumber
\\
a'+1 = \alpha +1- \gamma ,\;
b'+1 =  \beta+1- \gamma ,\;
c'+1 = 2- \gamma \; ,
\label{5.21a}
\end{eqnarray}

\noindent
we can expect
\begin{eqnarray}
2 \sqrt{z(1-z)}  \left ( {d \over dz} + {i\epsilon /2 \over 1-z} \right )
G^{(1)}_{non-zero}  + (M-\epsilon - i/2  ) \; F^{(2)}_{zero}  = 0 \;;
\label{5.21b}
\end{eqnarray}

\noindent
from  (\ref{5.21b}) it follows
\begin{eqnarray}
2 {a'b' \over c'} G^{non-zero}_{0} +  ( M -\epsilon - i/2 )  \;F^{zero}_{0}  = 0 \qquad\Longrightarrow
\nonumber
\\
 a' \; G^{non-zero}_{0}  + i   c' \; F^{zero}_{0}=0  \; .
\end{eqnarray}


\section{ Behavior of the solutions at the horizon,\\  standing and running waves, $j > j_{\min}$  }

First, consider a couple of linearly independent solutions $F^{(1)}_{reg}$ and  $F^{(1)}_{sing}$.
It is convenient to introduce  two other solutions
 with the help of the known Kummer's relation
 \begin{eqnarray}
U_{1} = {\Gamma(c) \Gamma (c-a-b) \over \Gamma (c-a) \Gamma (c-b)} \; U_{2} +
{\Gamma(c) \Gamma (a+b-c) \over \Gamma (a) \Gamma (b)} \; U_{6}  \; ,
\nonumber
\\
U_{5} = {\Gamma(2-c) \Gamma (c-a-b) \over \Gamma (1-a) \Gamma (1-b)} \; U_{2} +
{\Gamma(2-c) \Gamma (a+b-c) \over \Gamma (a+1-c) \Gamma (b+1-c)} \; U_{6}  \; ,
\nonumber
\\
\label{6.1}
\end{eqnarray}

\noindent
and  inverse ones
\begin{eqnarray}
U_{2} = {\Gamma(a+b+1- c) \Gamma (1-c) \over \Gamma (a+1-c) \Gamma (b+1-c )} \; U_{1} +
{\Gamma(a+b+1-c) \Gamma (c-1) \over \Gamma (a) \Gamma (b)} \; U_{5}  \; ,
\nonumber
\\
U_{6} = {\Gamma(c+1-a-b) \Gamma (1-c) \over \Gamma (1-a) \Gamma (1-b)} \; U_{1} +
{\Gamma(c+1-a-b) \Gamma (c-1) \over \Gamma (c-a) \Gamma (c-b)} \; U_{5}  \; ,
\nonumber
\\
\label{6.2}
\end{eqnarray}

\noindent where two couples of linearly independent solutions are involved:
\begin{eqnarray}
U_{1} (z) = F(a,b,c; z)  \;,
\nonumber
\\
 U_{5} = z^{1-c} F(a+1-c, b+1-c, 2-c, z )   \; ;
\nonumber
\\[5mm]
U_{2}(z) = F(a,b, a+b -c+1 ; 1-z)\;,
\nonumber
\\
U_{6} (z) = (1-z) ^{c-a-b} F(c-a, c-b, c-a-b +1; 1-z) \; .
\label{6.3}
\end{eqnarray}

Applying  relation (\ref{6.1}) to  the wave
$$
F^{(1)}_{reg}(z)= F^{(2)}_{reg}(z)= F_{reg} = z^{(\nu+1)/2} (1-z)^{-i\epsilon/2} U_{1} \; ,
$$

\noindent
 we obtain
\begin{eqnarray}
F_{reg}(z) =   z^{(\nu+1)/2} (1-z)^{-i\epsilon/2}  U_{1} =
z^{(1+\nu)/2} (1-z)^{-i\epsilon/2}
\nonumber
\\
\times\left \{
  {\Gamma(c) \Gamma (c-a-b) \over \Gamma (c-a) \Gamma (c-b)} \right.
      F(a,b, a+b -c +1 ; 1-z)
      \nonumber
  \\
 \left.
  + {\Gamma(c) \Gamma (a+b-c) \over \Gamma (a) \Gamma (b)}
(1-z) ^{+ i\epsilon +1/2} F(c-a, c-b, c-a-b +1; 1-z)  \right \}  ,
\nonumber
\end{eqnarray}

\noindent
so that
\begin{eqnarray}
F_{reg}(z)
= {\Gamma(c) \Gamma (c-a-b) \over \Gamma (c-a) \Gamma (c-b)} \; F_{out}   +
{\Gamma(c) \Gamma (a+b-c) \over \Gamma (a) \Gamma (b)}\;  F_{in} \; .
\label{6.3}
\end{eqnarray}

\noindent
Here  two independent solutions  with simple behavior on the horizon are noted according to
\begin{eqnarray}
F_{out} = z^{(\nu+1)/2} (1-z)^{-i\epsilon/2} U_{2} \; , \qquad
F_{in} =
z^{(\nu+1)/2} (1-z)^{-i\epsilon/2} U_{6}
\nonumber
\\
= z^{(\nu+1)/2}  (1-z) ^{(+ i\epsilon +1)/2} F(c-a, c-b, c-a-b +1; 1-z) \; .
\label{6.4}
\end{eqnarray}

Doing the same  for singular solutions
$$
F^{(1)}_{sing} = F^{(2)}_{sing}  = F_{sing} = z^{(\nu+1)/2} (1-z)^{-i\epsilon/2}  U_{5}\; ,
$$

\noindent we get
\begin{eqnarray}
F_{sing} = z^{(\nu+1)/2} (1-z)^{-i\epsilon/2} U_{5} = z^{(\nu+1)/2} (1-z)^{-i\epsilon/2}
\nonumber
\\
\times
 \left \{
{\Gamma(2-c) \Gamma (c-a-b) \over \Gamma (1-a) \Gamma (1-b)}
F(a,b, a+b -c +1 ; 1-z)   \right.
\nonumber
\\
\left.   + {\Gamma(2-c) \Gamma (a+b-c) \over \Gamma (a+1-c) \Gamma (b+1-c)}
(1-z) ^{+ i\epsilon +1/2}  \right.
\nonumber
\\
\left.
\times F(c-a, c-b, c-a-b +1; 1-z)  \right \} \; ,
\nonumber
\end{eqnarray}

\noindent
so that
\begin{eqnarray}
F_{sing}
= {\Gamma(2-c) \Gamma (c-a-b) \over \Gamma (1-a) \Gamma (1-b)} \; F_{out} +
 {\Gamma(2-c) \Gamma (a+b-c) \over \Gamma (a+1-c) \Gamma (b+1-c)} \;  F_{in} \; .
\label{6.5}
\end{eqnarray}

In a similar way let us consider solutions $G^{(1)}_{reg}(z)= G^{(2)}_{reg}(z)=G_{reg}$. Note that in relevant
Kummer's formulas we use $V$ instead of $U$.  Thus we get
\begin{eqnarray}
G_{reg}(z) =   z^{\nu/2} (1-z)^{+i\epsilon/2}  V_{1} =
z^{+\nu/2} (1-z)^{+i\epsilon/2}
\nonumber
\\
\times\left \{
  {\Gamma(c') \Gamma (c'-a'-b') \over \Gamma (c'-a') \Gamma (c'-b')} \right.
      F(a',b', a'+b' -c' +1 ; 1-z)
      \nonumber
  \\
 \left. +
  {\Gamma(c') \Gamma (a'+b'-c') \over \Gamma (a') \Gamma (b')}
(1-z) ^{- i\epsilon +1/2}   \right.
\nonumber
\\
\left.  \times F(c'-a', c'-b', c'-a'-b' +1; 1-z)  \right \}  \qquad \Longrightarrow
\nonumber
\end{eqnarray}
\begin{eqnarray}
G_{reg}(z) = {\Gamma(c') \Gamma (c'-a'-b') \over \Gamma (c'-a') \Gamma (c'-b')} \; G_{in} +
{\Gamma(c') \Gamma (a'+b'-c') \over \Gamma (a') \Gamma (b')} \; G_{out} \; ,
\nonumber
\\
\label{6.6}
\end{eqnarray}

\noindent where
\begin{eqnarray}
G_{in} = z^{\nu/2} (1-z)^{+i\epsilon/2} V_{2}
\nonumber
\\
= z^{\nu/2} (1-z)^{+i\epsilon/2}
z^{\nu/2} (1-z)^{+i\epsilon/2} F(a',b', a'+b' -c' +1 ; 1-z) \; ,
\nonumber
\\[4mm]
G_{out} =   z^{\nu/2} (1-z)^{+i\epsilon/2}  V_{6}
\nonumber
\\
=
z^{\nu/2}
(1-z) ^{(- i\epsilon +1) /2}    F(c'-a', c'-b', c'-a'-b' +1; 1-z) \; .
\nonumber
\\
\end{eqnarray}

\noindent For  singular ones, $G^{(1)}_{sing} = G^{(2)}_{sing} = G_{sing}$
\begin{eqnarray}
G^{(1)}_{sing} = z^{\nu /2} (1-z)^{+i\epsilon/2} V_{5} = z^{\nu/2} (1-z)^{+i\epsilon/2}
\nonumber
\\
\times
 \left \{
{\Gamma(2-c') \Gamma (c'-a'-b') \over \Gamma (1-a') \Gamma (1-b')}
F(a',b', a'+b' -c' +1 ; 1-z)   \right.
\nonumber
\\
 +  {\Gamma(2-c') \Gamma (a'+b'-c') \over \Gamma (a'+1-c') \Gamma (b'+1-c')}
\nonumber
\\
\left. \times
(1-z) ^{- i\epsilon +1/2} F(c'-a', c'-b', c'-a'-b' +1; 1-z)  \right \}
\nonumber
\end{eqnarray}

\noindent so that
\begin{eqnarray}
G^{(1)}_{sing} =
{\Gamma(2-c') \Gamma (c'-a'-b') \over \Gamma (1-a') \Gamma (1-b')} G_{in} +
  {\Gamma(2-c') \Gamma (a'+b'-c') \over \Gamma (a'+1-c') \Gamma (b'+1-c')} G_{out}
  \;.
  \nonumber
  \\
  \label{6.7}
  \end{eqnarray}

It should be mentioned that  the factors $(1-z)^{\pm i\epsilon /2}$ can be presented  like
plane waves. Indeed, let a new variable $x$  be
\begin{eqnarray}
1-z = e^{-2x}, \qquad
x = -{1 \over 2} \ln (1-z), \qquad x \in [0, + \infty ) \; ,
\nonumber
\\
\mbox{then} \qquad
(1-z)^{-i\epsilon /2} =  e^{i\epsilon x} \;,\qquad
(1-z)^{+i\epsilon /2} =  e^{-i\epsilon x} \;,\qquad  x \rightarrow + \infty \; .
\nonumber
\\
\label{plane}
\end{eqnarray}

Evidently the {\em out-}  and {\em in-}waves can be presented as linear combinations
 of {\em reg-} and {\em sing-}waves.
Relevant relations are
\begin{eqnarray}
F_{out} = z^{(\nu+1)/2} (1-z)^{-i\epsilon /2} U_{2} =
 z^{(\nu+1)/2} (1-z)^{-i\epsilon /2}
\nonumber
\\
\times
\left (
{\Gamma(a+b+1- c) \Gamma (1-c) \over \Gamma (a+1-c) \Gamma (b+1-c )} \; U_{1} +
{\Gamma(a+b+1-c) \Gamma (c-1) \over \Gamma (a) \Gamma (b)} \; U_{5}  \right )
\nonumber
\\
=
{\Gamma(a+b+1- c) \Gamma (1-c) \over \Gamma (a+1-c) \Gamma (b+1-c )} \; F_{reg}  +
{\Gamma(a+b+1-c) \Gamma (c-1) \over \Gamma (a) \Gamma (b)} \; F_{sing} \; ,
\nonumber
\\
\end{eqnarray}
\begin{eqnarray}
F_{in} = z^{(\nu+1)/2} (1-z)^{-i\epsilon /2} U_{6} = z^{(\nu+1)/2} (1-z)^{-i\epsilon /2}
\nonumber
\\
\times \left (
{\Gamma(c+1-a-b) \Gamma (1-c) \over \Gamma (1-a) \Gamma (1-b)} \; U_{1} +
{\Gamma(c+1-a-b) \Gamma (c-1) \over \Gamma (c-a) \Gamma (c-b)} \; U_{5}   \right )
\nonumber
\\
= {\Gamma(c+1-a-b) \Gamma (1-c) \over \Gamma (1-a) \Gamma (1-b)} \; F_{reg}  +
{\Gamma(c+1-a-b) \Gamma (c-1) \over \Gamma (c-a) \Gamma (c-b)} \; F_{sing} \;,
\nonumber
\\
\end{eqnarray}
\begin{eqnarray}
G_{in} = z^{\nu/2} (1-z)^{+i\epsilon/2} V_{2} = z^{\nu/2} (1-z)^{+i\epsilon/2}
\nonumber
\\
\times  \left (
{\Gamma(a'+b'+1- c') \Gamma (1-c') \over \Gamma (a'+1-c') \Gamma (b'+1-c' )} \; V_{1} +
{\Gamma(a'+b'+1-c') \Gamma (c'-1) \over \Gamma (a') \Gamma (b')} \; V_{5} \right )
\nonumber
\\
={\Gamma(a'+b'+1- c') \Gamma (1-c') \over \Gamma (a'+1-c') \Gamma (b'+1-c' )} \; G_{reg} +
{\Gamma(a'+b'+1-c') \Gamma (c'-1) \over \Gamma (a') \Gamma (b')} \; G_{sing} \; ,
\nonumber
\\
\end{eqnarray}
\begin{eqnarray}
G_{out} =   z^{\nu/2} (1-z)^{+i\epsilon/2}  V_{6} = z^{\nu/2} (1-z)^{+i\epsilon/2}
\nonumber
\\
\times \left (  {\Gamma(c'+1-a'-b') \Gamma (1-c') \over \Gamma (1-a') \Gamma (1-b')} \; V_{1} +
{\Gamma(c'+1-a'-b') \Gamma (c'-1) \over \Gamma (c'-a') \Gamma (c'-b')} \; V_{5} \right )
\nonumber
\\
= {\Gamma(c'+1-a'-b') \Gamma (1-c') \over \Gamma (1-a') \Gamma (1-b')} \; G_{reg} +
{\Gamma(c'+1-a'-b') \Gamma (c'-1) \over \Gamma (c'-a') \Gamma (c'-b')} \; G_{sing} \; .
\nonumber
\\
\end{eqnarray}

\section{ Standing and running waves at   $j=j_{\min}$  }

Let write down   results we need to proceed further
\begin{eqnarray}
F_{non-zero} =   (1-z)^{-i\epsilon/2}  U_{1}\; , \qquad
F_{zero} = (1-z)^{-i\epsilon/2}  U_{5} \; ,
\nonumber
\\
G_{non-zero} =   (1-z)^{+i\epsilon/2}  V_{1}\; , \qquad
G_{zero} = (1-z)^{+i\epsilon/2}  V_{5} \; ,
\label{7.1}
\end{eqnarray}
\begin{eqnarray}
U_{1} = F(a,b,c, z)\; , \qquad V_{1} = F(a',b',c', z)\; ,
\nonumber
\\
a = {-i\epsilon \over 2} + {iM +1/2 \over 2}\;, \qquad
b= {-i\epsilon \over 2} - {iM +1/2 \over 2}\;, \qquad c  = 1/2\; ,
\nonumber
\\
a' = {+i\epsilon \over 2} + {iM +1/2 \over 2}\;, \qquad
b'= {+i\epsilon \over 2} - {iM +1/2 \over 2}\;, \qquad c'  = 1/2 \; .
\label{7.2}
\end{eqnarray}

We readily derive
\begin{eqnarray}
F_{non-zero} =   (1-z)^{-i\epsilon/2}  U_{1}
\nonumber
\\
=
{\Gamma(c) \Gamma (c-a-b) \over \Gamma (c-a) \Gamma (c-b)} \; F_{out} +
{\Gamma(c) \Gamma (a+b-c) \over \Gamma (a) \Gamma (b)} \; F_{in}\; ,
\nonumber
\\
F_{zero} =   (1-z)^{-i\epsilon/2}  U_{5}
\nonumber
\\
=
{\Gamma(2-c) \Gamma (c-a-b) \over \Gamma (1-a) \Gamma (1-b)}  F_{out} +
{\Gamma(2-c) \Gamma (a+b-c) \over \Gamma (a+1-c) \Gamma (b+1-c)}  \; F_{in}\; ,
\nonumber
\\
F_{out}= (1-z)^{-i\epsilon/2} U_{2} \; , \qquad  F_{in}= (1-z)^{-i\epsilon/2} U_{6} \; .
\end{eqnarray}
\begin{eqnarray}
G_{non-zero} =   (1-z)^{+i\epsilon/2}  V_{1}
\nonumber
\\
=
{\Gamma(c) \Gamma (c-a-b) \over \Gamma (c-a) \Gamma (c-b)} \; G_{in} +
{\Gamma(c) \Gamma (a+b-c) \over \Gamma (a) \Gamma (b)} \; G_{out}\; ,
\nonumber
\\
G_{zero} =   (1-z)^{+i\epsilon/2}  V_{5}
\nonumber
\\
=
{\Gamma(2-c') \Gamma (c'-a'-b') \over \Gamma (1-a') \Gamma (1-b')}  G_{in} +
{\Gamma(2-c') \Gamma (a'+b'-c') \over \Gamma (a'+1-c') \Gamma (b'+1-c')}  \; G_{out}\; ,
\nonumber
\\
G_{in}= (1-z)^{+i\epsilon/2} V_{2} \; , \qquad  G_{out}= (1-z)^{+i\epsilon/2} V_{6} \; .
\nonumber
\\
\end{eqnarray}

\section{ Discussion and  conclusions }

To understand  better the situation, let us consider the case of
 minimal $j_{\min}$  in
 the limit of vanishing curvature.
It is convenient to start with   the  first order systems for minimal values $j_{\min}$
in the case of Minkowski space:

\vspace{3mm}
$
k = +1/2,+1,\ldots $
\begin{eqnarray}
\epsilon  \; f_{3} - i \;  { d\over dr}  \; f_{3}  - M \; f_{1} = 0\; ,
\nonumber
\\
 \epsilon  \; f_{1} + i \;  { d \over dr} \; f_{1}  - M \; f_{3} = 0 \; ;
\label{C.5}
\end{eqnarray}

$
k = -1/2,-1,\ldots $
\begin{eqnarray}
\epsilon   \; f_{4} + i \;  { d\over dr}\; f_{4} - M \;f_{2} = 0 \; ,
\nonumber
\\
 \epsilon \; f_{2} - i \;  { d\over dr}\; f_{2} - M \;f_{4} = 0 \; .
\label{C.6}
\end{eqnarray}

Let  us detail the case of positive $
k = +1/2,+1,\ldots $. With notation
\begin{eqnarray}
{f_{1} + f_{3} \over \sqrt{2} } = h(r) \; , \qquad  { f_{1} - f_{3} \over  i \sqrt{2} } = g(r) \; ;
\label{C.7}
\end{eqnarray}

\noindent
relevant equations are
\begin{eqnarray}
{d \over dr }h +( \epsilon + M) \; g = 0 \; , \qquad
{d \over dr} g - (\epsilon - M) h = 0 \; .
\label{C.8}
\end{eqnarray}

\noindent Further, with the substitutions
\begin{eqnarray}
h(r) = H e^{\gamma r}\;, \qquad g (r) = G e^{\gamma r}
\label{C.9}
\end{eqnarray}

\noindent
we get (first let it be $(\epsilon^{2} - M^{2}) > 0 $)
\begin{eqnarray}
\gamma^{2} = - (\epsilon^{2} - M^{2})  = -  p ^{2}  \;,
\nonumber
\\
 \gamma = +i p, - i p \; , \qquad G \gamma - (\epsilon - M) H = 0
\label{C.10}
\end{eqnarray}

\noindent
Thus we have two linearly independent solutions
\begin{eqnarray}
h_{1}(r) = H_{1} e^{+ip r} \; , \qquad g_{1}(r) = G_{1} e^{+ipr} \; ,
\qquad
G_{1} = {\epsilon - M \over  ip} H_{1} \; ;
\label{C.11a}
\end{eqnarray}
and
\begin{eqnarray}
h_{2}(r) = H_{2} e^{-ip r} \; , \qquad g_{2}(r) = G_{2} e^{-ipr} \; ,
\qquad
 G_{2} = {\epsilon - M \over  - ip} H_{2} \; ;
\label{C.11b}
\end{eqnarray}

\noindent
for simplicity, we  will take $H_{1}=H_{2} = 1$.
It is convenient to  introduce  linear combinations of these solutions

\vspace{3mm}

\underline{the first}
\begin{eqnarray}
{ h_{1}(r) + h_{2}(r) \over 2 }= \cos pr \;,
\nonumber
\\
{ g_{1}(r) + g_{2}(r) \over 2 }
=
{\epsilon - M \over  p} \;  \sin pr \; ;
\label{C.12}
\end{eqnarray}

\underline{the second}
\begin{eqnarray}
{ h_{1}(r) - h_{2}(r) \over 2 i }= \sin pr \;,
\nonumber
\\
{ g_{1}(r) - g_{2}(r) \over 2 i }=
{\epsilon - M \over - p} \;  \cos pr \; .
\label{C.13}
\end{eqnarray}

Now let us specify the case
 $(\epsilon^{2} - M^{2}) < 0 $:
\begin{eqnarray}
\gamma^{2} = - (\epsilon^{2} - M^{2}) \equiv = +  q ^{2}  \;, \qquad \gamma = + q, - q \; ,
\nonumber
\\
G \gamma - (\epsilon - M) H = 0 \; .
\label{C.14}
\end{eqnarray}

\noindent We  have two linearly independent solutions
\begin{eqnarray}
h_{1}(r) = H_{1} e^{+q r} \; , \qquad g_{1}(r) = G_{1} e^{+qr} \; ,
\qquad G_{1} = {\epsilon - M \over  q} H_{1} \; ;
\label{C.15a}
\end{eqnarray}
\begin{eqnarray}
h_{2}(r) = H_{2} e^{-q r} \; , \qquad g_{2}(r) = G_{2} e^{-q r} \; , \qquad G_{2} = {\epsilon - M \over  - q} H_{2} \; .
\label{C.15b}
\end{eqnarray}

\noindent
Below,  $H_{1}=H_{2} = 1$. We can introduce two linear combinations of these solutions

\vspace{3mm}
\underline{the first}
\begin{eqnarray}
{ h_{1}(r) + h_{2}(r) \over 2 }= \mbox{cosh}\;  qr \;,
\nonumber
\\
{ g_{1}(r) + g_{2}(r) \over 2 }=
{\epsilon - M \over  q} \;  \mbox{sinh}\;  qr
\label{C.16}
\end{eqnarray}

\underline{the second}
\begin{eqnarray}
{ h_{1}(r) - h_{2}(r) \over 2  }=  \mbox{sinh}\;  qr \;,
\nonumber
\\
{ g_{1}(r) - g_{2}(r) \over 2  }=
{\epsilon - M \over  q} \;  \mbox{cosh}\;  qr \; .
\label{C.17}
\end{eqnarray}

Evidently, above  constructed  solutions in de Sitter model provide us with  generalizations of
these in  Minkowski space.
 It may be verified additionally by
direct limiting process when $\rho \rightarrow \infty$.
To this end, let us translate  solutions in de Sitter space to usual units $\rho$ is the curvature radius,
$E$ is the energy, $c$ is the light velocity, $m$ is the electron mass)
\begin{eqnarray}
F_{non-zero}(R)  =   \left ( 1 -{R^{2}\over \rho^{2} } \right )^{-i{E\rho \over  2c \hbar }}
 F(a,b,c;  {R^{2} \over \rho^{2}}) \;,
\nonumber
\\
F_{zero}(R)  = R     \left ( 1 -{R^{2}\over \rho^{2} } \right )^{+i{E\rho \over  2c \hbar} +1/2 }
F(a+1-c,b+1-c,2-c; {R^{2} \over \rho^{2}})\; ,
\nonumber
\end{eqnarray}
\begin{eqnarray}
G_{non-zero} (R)  =  \left ( 1 -{R^{2}\over \rho^{2} } \right )^{+i{E\rho \over 2c \hbar} }
F(a',b',c; {R^{2} \over \rho^{2}})  \; ,
\nonumber
\end{eqnarray}
\begin{eqnarray}
G_{zero}(R)  =  R    \left ( 1 -{R^{2}\over \rho^{2} } \right )
^{-i{E\rho \over 2 c \hbar} +1/2 }   F(a'+1-c,b'+1-c,2-c; {R^{2} \over \rho^{2}})  \; ,
\nonumber
\end{eqnarray}

\noindent
Parameters of hypergeometric functions are given by
\begin{eqnarray}
c = {1 \over 2} \;  , \;
a = {1 \over 2} \left  [  +1/2  + i ( {mc\rho \over \hbar} - {E \rho \over c \hbar } ) \right   ]  \;, \;
b =   {1 \over 2} \left[   -i ({mc\rho \over \hbar}+ {E \rho \over c \hbar } ) - 1/2 \right  ] \; ,
\nonumber
\end{eqnarray}

\begin{eqnarray}
c = {1 \over 2} \;  , \;
a' = {1 \over 2} \left[ +1/2  + i ( {mc\rho \over \hbar} +  {E \rho \over c \hbar } ) \right ]  \;, \;
b '=    {1 \over 2}\left [   -i({mc\rho \over \hbar}- {E \rho \over c \hbar }) - 1/2  \right] \;.
\nonumber
\end{eqnarray}

Let us examine the limiting procedure  at $\rho \rightarrow \infty$ in
$
 F(a,b,c;  R^{2} /  \rho^{2}) \; .
$
Because
\begin{eqnarray}
{1 \over 1!} \; {ab \over c}  \;  {R^{2} \over \rho^{2}}\;\;  \rightarrow  \;\;  {1 \over 2!}  ( m^{2}c^{2} /
 \hbar^{2} -E^{2} / \hbar^{2} c^{2} )R^{2}  = - {1 \over 2!} \; (pR)^{2}\; ,
\nonumber
\\
{1 \over 2!} \; {a(a+1)b (b+1)  \over c(c+1)} \;  {R^{2} \over \rho^{2}}\;\;  \rightarrow
 \;\;    + { (pR)^{4} \over  4! } \; ,
\nonumber
\\
{1 \over 3!} \; {a(a+1)(a+2)b (b+1) (b+2) \over c(c+1)(c+2)} \;  {R^{2} \over \rho^{2}}\;\;  \rightarrow
 \;\;    - { (pR)^{6} \over  6! } \; ,
\nonumber
\end{eqnarray}

\noindent and so on, we obtain  the following limiting relation
\begin{eqnarray}
\lim_{\rho \infty }  F(a,b,c;  {R^{2} \over \rho^{2}})  = \cos pr \qquad \Longrightarrow
\nonumber
\\
\lim_{\rho \infty }  F_{non-zero} (R)   = \cos pr \; , \qquad
\lim_{\rho \infty }  G_{non-zero} (R)   = \cos pr \; .
\end{eqnarray}

In the same manner, let us examine  the function
\begin{eqnarray}
R     \;    F(a+1-c,b+1-c,2-c; {R^{2} \over \rho^{2}}) \; ,
\nonumber
\\
A = a+1-c ={3/2 +i(M + \epsilon ) \over 2} \;  ,
\nonumber
\\
B = b+1-c = {1/2 -i(M - \epsilon ) \over 2}, \qquad C = 3/2 \; .
\nonumber
\end{eqnarray}

\noindent
Taking into account
\begin{eqnarray}
{AB \over C} \;\; \Longrightarrow\; \; - {1 \over  3!}\; (pR) ^{2} \; ,\qquad
{1 \over 2!} \; {A(A+1)B(B+1)  \over C(C+1) } \;\; \Longrightarrow\; \; + {1 \over  5!}\; (pR) ^{4} \; ,
\nonumber
\\
{1 \over 3!} \; {A(A+1)(A+2)B(B+1)(B+2)  \over C(C+1)(C+2)  } \;\; \Longrightarrow\; \; - {1 \over  7!}\; (pR) ^{6} \; ,
\nonumber
\end{eqnarray}

\noindent and so on,  we arrive at the relationships
\begin{eqnarray}
\lim _{\rho \infty } \; p R     \; \;    F(a+1-c,b+1-c,2-c; {R^{2} \over \rho^{2}})  = \sin p R \; \qquad
\Longrightarrow
\nonumber
\\
\lim _{\rho \infty } \; p R      \; F_{zero}  = \sin p R \;, \qquad
\lim _{\rho \infty } \; p R      \; G_{zero}  = \sin p R \;.
\end{eqnarray}

Thus, indeed, solutions in de Sitter model are extensions of more simple  and well-known ones in Minkowski model.

\section{Acknowledgements}

This  work was   supported   by the Fund for Basic Researches of Belarus
 F11M-152.

\end{document}